\documentstyle[aps]{revtex}
\begin{document}
\title{Non exponential relaxation in fully frustrated models}
\author{Annalisa Fierro$^1$, Antonio de Candia$^{1,2}$,
        Antonio Coniglio$^{1,2}$}
\address{$^1$Dipartimento di Fisica,
         Mostra d'Oltremare, pad. 19, 80125 Napoli, Italy}
\address{$^2$INFM, Sezione di Napoli, Napoli, Italy}
\maketitle
\begin{abstract}
We study the dynamical properties of the fully frustrated Ising model.
Due to the absence of disorder the model, contrary to spin glass, does
not exhibit any Griffiths phase, which has been associated to
non-exponential relaxation dynamics.
Nevertheless we find numerically that the model
exhibits a stretched exponential
behavior below a temperature $T_p$ corresponding to the percolation transition
of the Kasteleyn-Fortuin clusters.
We have also found that the critical behavior of this clusters for a
fully frustrated
$q$-state spin model at the percolation threshold is strongly affected by
frustration. In fact while in absence of frustration the $q=1$ limit
gives random percolation, in presence of frustration the critical behavior
is in the same universality class of the ferromagnetic $q=1/2$-state Potts
model.
\end{abstract}
\pacs{}
%
%%%%%%%%%%%%%%%%%%%%%%%%%%%%%%%%%%%%%%%%%%%%%%%%%%%%%%%%%%%%%%%%%%%%%
%
\section{Introduction}
Glassy systems at low temperatures undergo a transition characterized
by the freezing of structural relaxation, in which the system is
trapped in a disordered metastable configuration.
Already at temperatures higher than the ideal glass transition temperature
$T_0$, a number of dynamical anomalies is observed.
One of these anomalies concerns the relaxation functions of the system, which
at  high temperatures are characterized by a single exponential. Below
some temperature $T^\ast$ higher than $T_0$, the long time regime of
correlation functions, called $\alpha$-relaxation, is well  approximated
by a Kohlrausch-Williams-Watts function, also known as
``stretched exponential'',
\begin{equation}
f(t)=f_0\exp-(t/\tau)^\beta
\label{eq_stretched}
\end{equation}
This behavior has been observed in many real glasses,
such as ionic conductors, supercooled liquids and polymers
\cite{ref_johari,ref_william,ref_ngaiuno,ref_strom,%
ref_angell,ref_ngaidue,ref_weiler}.
\par
A similar behavior has been observed in canonical metallic and
insulating spin glasses,
investigated by neutron and hyperfine techniques
\cite{ref_mezei,ref_murani,ref_mezeidue,ref_huser,ref_uemera,ref_meyer}.
These systems can be described by an Ising model, in which ferromagnetic and
antiferromagnetic interactions are distributed in a disordered way on the
edges of the lattice.
The Ising spin glass undergoes a transition at some temperature $T_{sg}$,
called spin glass transition,
analogous to the freezing transition of real glasses.
Moreover, like in glass-forming
systems, one observes a temperature value $T^\ast>T_{sg}$ where
non-exponential relaxation functions appear. This has been observed for
the Ising spin glass in 2D by McMillan \cite{ref_mcmillan}, and in 3D
by Ogielski \cite{ref_ogielski}.
\par
Several mechanisms have been proposed to explain the onset of
non exponential relaxation functions like (\ref{eq_stretched}) in spin
glasses, when the system approaches the glass transition from above.
Randeria {\em et al.} \cite{ref_randeria} suggest that the
temperature $T^\ast$
coincides with $T_c$, the critical temperature of the ferromagnetic model.
They base their conjecture on the presence, in the spin glass, of
non frustrated ferromagnetic-type
clusters of interactions, the same that are responsible for the
Griffiths singularity \cite{ref_griffiths}.
The presence of non exponential relaxation in this approach is therefore
a direct consequence of the quenched disorder.
\par
On the other hand,
it has been suggested  \cite{ref_campbell,ref_silvia,ref_glotzer}
that in the spin glass the onset
of stretched exponential relaxation functions may coincide
with the percolation
temperature $T_p$ of the Kasteleyn-Fortuin and Coniglio-Klein clusters
\cite{ref_kasteleyn,ref_cklein}.
These clusters can be obtained introducing bond with probability
$p_B=1-e^{-2J/k_BT}$ between nearest-neighbor pair of spins satisfying the
interaction.
In the spin glass the percolation threshold of these clusters
is higher than $T_{sg}$ \cite{ref_diliberto}.
Moreover Stauffer \cite{ref_stauffer} observes non exponential relaxation
in the $d=2$ ferromagnetic Ising model,
simulated by conventional spin flip, for temperatures lower than
the critical temperature $T_c$, which  coincides with
the percolation temperature $T_p$.
In $d=3$ this behavior disappears, and the relaxation is purely exponential
for all the temperatures.
\par
To gain more insight  into the mechanisms which lead
to the appearing of anomalies in the dynamical behavior, in this paper
we study
the fully frustrated Ising model \cite{ref_villain,ref_forgacs,ref_wolff}
on a bidimensional square lattice,
using the conventional spin flip techniques.
In this model ferromagnetic and antiferromagnetic interactions are distributed
in a regular way on the lattice (see Fig. \ref{fig_fully}),
so that no unfrustrated cluster of  interactions
exists.
Therefore in this model the appearing of stretched exponential in the
relaxation functions cannot be due to the mechanism of
Randeria {\em et al.}
We in fact find that the onset of stretched exponentials occurs
at the percolation temperature $T_p$, as we show in Sect. \ref{sec_sflip}.
\par
The fully frustrated Ising model can be mapped onto a fully frustrated
$q$-bond percolation by applying the Kasteleyn-Fortuin and Coniglio-Klein
cluster formalism, where $q=2$ is the multiplicity of the spins
(see Sect. \ref{sec_model}).
This model, which can be generalized to any value of $q$, is suitable
to describe systems with geometrical frustration, and may give
insight into the origin of the long relaxation decay in glasses,
characterized by stretched exponentials
\cite{ref_coniglio}.
We study the dynamics of this geometrical model for $q=1$ and $q=2$,
using the ``bond flip'' dynamics, that will be described in Sect.
\ref{sec_montecarlo}. We find again that the onset of stretched
exponentials starts at the percolation transition $T_p$ (Sect.
\ref{sec_relaxation}).
Moreover we also find that the percolation transition is in the same
universality class of the $q/2$-state ferromagnetic Potts model,
as we show in Sect. \ref{sec_statics},
in agreement with the expectation based on renormalization group
\cite{ref_pezzella}.
\par
These results thus suggest that the presence of a percolation type transition
may be responsible for the appearing of the ``large scale''
effects of frustration, among which there is the
onset of various dynamical anomalies, such as
stretched exponential relaxation functions \cite{ref_jan,ref_glotzer}.
\par
In the frustrated $q$-bond percolation, the frustration is present
at all length scale. To probe the effect of frustration we have modified
the model in such a way that only loops of length four are considered
frustrated
(Sect. \ref{sec_local}).
We find that the model for $q=1$
has the same critical exponents of the ferromagnetic Potts model with
spin multiplicity $q=1$ (the random bond percolation model).
Namely this local frustration does not change the critical behavior
compared with the unfrustrated model. At the same time,
although the dynamics of
the model is influenced by the frustration constraint,
in the long time regime the relaxation is purely exponential.
\section{The relaxation functions of the fully frustrated Ising spin model}
\label{sec_sflip}
We simulate by conventional spin flip the fully frustrated Ising spin model,
defined by the Hamiltonian
\begin{equation}
{\cal H}=-J\sum_{\langle{ij}\rangle}(\epsilon_{ij}S_{i}S_{j}-1),
\label{eq_Ising}
\end{equation}
where $\epsilon_{ij}$ are quenched variables which assume the
values $\pm 1$. The ferromagnetic and antiferromagnetic interactions are
distributed in a regular way on the lattice (see Fig. \ref{fig_fully}).
\par
We calculate the relaxation functions of the energy. Averages were made over
32 different random number generator seeds, and between $8\times 10^5$ and
$1.8\times 10^6$ steps for acquisition were taken, after $10^4$ steps for
thermalization, on a system of size $L=64$.
Here a unit of time is considered to be one Monte Carlo
step, that is $L^2$ single spin update trials.
\par
In Fig. \ref{fig_relax_sflip} we show the results for
$T=11.0,4.5,2.269,1.701,1.110$, where temperatures are expressed
in units of $J/k_B$. We observe
a two step decay also for very high temperatures. For all the temperatures
except $T=11.0$ we fit only the long time tail of the relaxation functions.
For temperatures $T>T_p=1.701$ we could fit the long time regime a pure
exponential, that is $\beta=1$ within the errors in Eq.
(\ref{eq_stretched_fit}).
On the other hand for $T=1.110<T_p$ the
long time behavior is not purely exponential, but can be fitted
asymptotically with a stretched exponential.
In Fig. \ref{fig_sflip_betas} we show the values of $\beta(T)$ in function
of $T/T_p$.
\section{The ``\lowercase{$q$}-bond frustrated percolation'' model}
\label{sec_model}
Using the Kasteleyn-Fortuin \cite{ref_kasteleyn} and
Coniglio-Klein \cite{ref_cklein} cluster formalism for
frustrated spin Hamiltonians \cite{ref_diliberto},
it is possible to show that the partition function of the model Hamiltonian
(\ref{eq_Ising})
is given by
\begin{equation}
Z={\sum_C}^\ast e^{\beta\mu n(C)} q^{N(C)},
\label{eq_partition}
\end{equation}
where $q=2$ is the multiplicity of the spins, $\beta=1/k_BT$,
$\mu=k_BT\ln(e^{q\beta J}-1)$,
$n(C)$ and $N(C)$ are respectively the number of bonds and the number of
clusters in the bond configuration $C$. The summation $\sum_C^\ast$ extends
over all the bond configurations that do not contain a ``frustrated loop'',
that is a closed path of bonds which contains  an odd number of
antiferromagnetic interactions.
Note that there is only one parameter in the model, namely the temperature
$T$, ranging from $0$ to $\infty$. The parameter $\mu$,
that can assume positive
or negative values, plays the role of a chemical potential.
\par
Varying $q$ we obtain an entire class of models differing by the
``multiplicity'' of the spins, which we call the
$q$-bond fully frustrated percolation model.
More precisely, for a general value of $q$,
the model can be obtained from a Hamiltonian \cite{ref_vittorio}
\begin{equation}
{\cal H}=-sJ\sum_{\langle{ij}\rangle}
[(\epsilon_{ij}S_{i}S_{j}+1)\delta_{\sigma_{i}\sigma_{j}}-2],
\end{equation}
in which every site carries two types of spin, namely an Ising spin and a Potts
spin $\sigma_i=1,\ldots,s$ with $s=q/2$.
For $q=1$ the factor $q^{N(C)}$ disappears
from Eq. (\ref{eq_partition}), and we obtain a simpler model in which
the bonds are randomly distributed under the conditions that the bond
configurations do not contain a frustrated loop.
For $q\to 0$ we recover the tree percolation,
in which all loops are forbidden, be they frustrated or not
\cite{ref_wu}.
\par
When all the interactions are positive (i.e. $\epsilon_{ij}=1$)
the sum in (\ref{eq_partition}) contains all bond configurations
without any restriction. In this case the partition function coincides
with the partition function of the ferromagnetic $q$-state Potts model,
which in the limit $q=1$ gives the random bond percolation.
\par
{}From renormalization group and numerical results we expect that the
model (\ref{eq_partition}) exhibits two critical points
\cite{ref_pezzella,ref_adc}.
The first at a temperature $T_p(q)$, corresponding to the percolation
of the bonds on the lattice, in the same universality class of the
ferromagnetic $q/2$-state Potts model.
The other at a lower temperature $T_{0}(q)$, in the
same universality class of the fully frustrated Ising model.
In the bidimensional case
$T_{0}(q)=0$ \cite{ref_villain,ref_forgacs,ref_wolff}.
\section{Monte Carlo dynamics}
\label{sec_montecarlo}
A particular configuration of the model defined by (\ref{eq_partition})
is determined by the state of each edge between two nearest-neighbor sites,
that can be empty or occupied by a bond. The dynamics of the model is
carried out in the following way: (i) choose at random a particular
edge on the lattice; (ii) calculate the probability $P$
of changing its state,
that is of creating a bond if the edge is empty, and of destroying the bond
if the edge is occupied; (iii)  change the state of the edge
with probability $P$.
\par
The point (ii) needs the knowledge of a non local property, namely if a bond
placed on the chosen edge
closes a loop or not, and if the loop is frustrated or not.
This is accomplished in the following way: starting from the two sites
at the extremes of
the edge, visit the clusters of sites connected to them by a
continuous path of bonds; if the clusters collide the bond closes a loop,
otherwise it does not.
Taking track of the number of antiferromagnetic bonds traversed visiting
the cluster, one can determine also if the loop is frustrated or not.
\par
Note that at high temperatures clusters are small,
and are visited in a few iterations, while at low temperatures
density of bonds  is high, and the clusters collide in a few iterations
as well.   On the other hand at the percolation transition clusters are
very ramified, and one often must visit a large number of sites  before the
iteration is over. This makes the algorithm CPU consuming at the percolation
transition, and prevents the simulation of very large systems.
\par
{}From Eq. (\ref{eq_partition}),
the statistical weight of a bond configuration $C$ is given  by
$W(C)=e^{\beta\mu n(C)}q^{N(C)}$ if $C$ does not contain a frustrated
loop, and $W(C)=0$ if it does.
Thus the transition probabilities for the principle
of detailed balance must satisfy
\begin{equation}
P(C\to C^\prime) = P(C^\prime\to C)e^{\beta\mu\delta n}q^{\delta N},
\end{equation}
where $\delta n=n(C^\prime)-n(C)$ and $\delta N=N(C^\prime)-N(C)$.
Note that from Euler's equation $\delta N=\delta\kappa-\delta n$,
where $\delta\kappa=\kappa(C^\prime)-\kappa(C)$, and $\kappa(C)$
is the number of loops in configuration $C$, we can calculate
$\delta N$ knowing $\delta n$ and $\delta\kappa$.
One can easily see that a possible choice for the transition
probability $P(C\to C^\prime)$ is given by
\begin{equation}
P(C\to C^\prime)=
\left\{\begin{array}{ll}
\min(1,e^{\beta\mu\delta n}q^{\delta N})\>&
\text{if $C^\prime$ is not frustrated,}\\
0&\text{if $C^\prime$ is frustrated.}
\end{array}\right.
\end{equation}
\par
The procedure described above in the points (i)---(iii) is called
a ``single update trial''.
A Monte Carlo step consists in ${\cal G}$ single update trials,
where ${\cal G}$ is the total number of edge on the lattice,
namely, on the square bidimensional lattice, ${\cal G}=2L^2$.
In Sect. \ref{sec_relaxation} and \ref{sec_local}, when we plot
relaxation functions of the fully frustrated and locally frustrated
bond percolation model, a unity of time is considered to be
${\cal G}\langle\rho\rangle^{-1}$ single update trials,
or $\langle\rho\rangle^{-1}$ Monte Carlo steps,
where $\langle\rho\rangle$ is the average density of bonds, ranging
in the interval $(0,1)$.
\section{Static properties}
\label{sec_statics}
In this Section we analyse the percolation properties of the model defined
by Eq. (\ref{eq_partition}), for $q=1$ and $q=2$, on a bidimensional square
lattice, with fully frustrated interactions.
We have used the histogram method for analyzing data
\cite{ref_swendsen}. For each value of $q$, we have simulated the model
for lattice sizes $L=32,48,64$. For each size we have considered $10$
temperatures around the percolation point, taking $10^4$ steps
for thermalization and between $3\times 10^5$ and $8\times 10^5$ steps
for acquisition of histograms. At every step we evaluate the following
quantities: density of bonds $\rho$; existence of a spanning cluster
$P_\infty$; mean cluster size $\chi$.
\par
The quantity   $P_\infty$ assumes the value $1$ if the bond configuration
percolates, and $0$ if it does not. The mean cluster size is defined as
\cite{ref_aharony}
\begin{equation}
\chi=\frac{1}{\cal N}\sum_{s} s^2n_s,
\end{equation}
where $n_s$ is the number of clusters having size $s$ on the lattice,
and ${\cal N}=L^2$ is the number of sites.
The histogram method allows to evaluate the thermal averages of this quantities
over an entire interval of temperature. The average of the quantity
$P_\infty$ is the probability of occurrence of a spanning cluster.
\par
Around the percolation temperature, the averaged
quantities $P_\infty(T)$ and
$\chi(T)$, for different values of the lattice size $L$,
should obey the finite size scaling \cite{ref_binder}
\begin{mathletters}
\begin{equation}
P_\infty(T) = F_\infty[L^{1/\nu}(T-T_p)],
\end{equation}
\begin{equation}
\chi(T) = L^{\gamma/\nu}F_\chi[L^{1/\nu}(T-T_p)],
\end{equation}
\end{mathletters}
where $\gamma$ and $\nu$ are critical exponents of mean cluster size
and connectivity length, $F_\infty$ and $F_\chi$ are universal functions.
Thus we can fit the values of $\nu$, $\gamma$ and $T_p$ so that plotting
$P_\infty(T)$ and $L^{-\gamma/\nu}\chi(T)$ in function of
$L^{1/\nu}(T-T_p)$, the functions corresponding to different values
of $L$ collapse respectively on the universal master curves
$y=F_\infty[x]$ and $y=F_\chi[x]$.
Figures \ref{fig_statics_q1} and \ref{fig_statics_q2} show the data
collapse obtained for the $q=1$ and $q=2$ models, for lattice sizes
$L=32,48,64$.
\par
The values of the critical exponents extracted
from the fit coincides, within the errors, with those of the
ferromagnetic Potts model with spin multiplicity $q/2$ \cite{ref_wu}.
Results are summarized in Tab. \ref{tab_critical},
while in Tab. \ref{tab_potts} we report the critical exponents and
transition temperature of the ferromagnetic Potts model.
\section{The relaxation functions of the
``\lowercase{$q$}-bond frustrated percolation'' model}
\label{sec_relaxation}
In this Section we analyse the dynamical behavior of the model defined by
(\ref{eq_partition}), simulated by the algorithm described in
Sect. \ref{sec_montecarlo}. For each temperature $T$ and value of $q$,
32 different runs were made, varying the random number generator seed,
on a system of size $L=64$. We took between $10^3$ and $10^4$ steps
for thermalization, and between $10^5$ and $10^6$ steps
for acquisition, calculating at each step the density of bonds $\rho(t)$.
The relaxation function of the density of bonds is defined as
\begin{equation}
f(t)=\frac{\langle\delta\rho(t)\delta\rho(0)\rangle}
{\langle(\delta\rho)^2\rangle},
\end{equation}
where $\delta\rho(t)=\rho(t)-\langle\rho\rangle$. For each value of
$T$ and $q$, we averaged the 32 functions calculated, and evaluated the
error as a standard deviation of the mean.
As we mentioned in Sect. \ref{sec_montecarlo}, we consider a unit of time
to consist of ${\cal G}\langle\rho\rangle^{-1}$ single update trials,
where ${\cal G}=2L^2$ is the number of edges on the lattice.
\par
In Fig. \ref{fig_relax_1} the results for $q=1$,
$T=1.440,1.067,0.801,0.625$ are shown.
Note that $T_p=1.067$ corresponds
to the percolation transition of the model.
For $T>T_p$ we fitted the calculated points with the function
\begin{equation}
f(t)=f_0\exp-(t/\tau)^\beta.
\label{eq_stretched_fit}
\end{equation}
The value of $\beta$ extracted from the fit is equal to one
within the error. Thus for $T\geq T_p$ the relaxation is purely exponential.
For $T<T_p$, we observe a two step decay, and only the long time regime of the
relaxation functions could be fitted by Eq. (\ref{eq_stretched_fit}).
The value of $\beta$ extracted is less than one, showing that stretched
exponential relaxation has appeared for these temperatures.
In Fig. \ref{fig_beta_1} the values of $\beta(T)$ in function of the
the ratio $T/T_p$ are shown, with least squares estimation errors.
\par
In Fig. \ref{fig_relax_2} the results for $q=2$,
$T=2.269,1.701,1.440,1.110$ are
shown. Temperature $T_p=1.701$ corresponds to the percolation transition.
The fits were made in the same way described for $q=1$, and the values of
$\beta(T)$ extracted are shown in Fig. \ref{fig_beta_2}.
Also in this case $\beta=1$ within the error for $T\geq T_p$, and
$\beta<1$ for $T<T_p$.
Note that
the $q=2$ fully frustrated bond percolation model can be mapped exactly
onto the fully frustrated Ising spin model, as we showed in Sect.
\ref{sec_model}. So it is interesting to compare the relaxation functions
of this model to those of the corresponding  fully frustrated Ising model,
simulated by standard spin flip techniques.
\section{The locally frustrated bond percolation}
\label{sec_local}
In the fully frustrated $q$-bond percolation model, the
configurations of bonds  which contain at least one frustrated loop
have zero weight. The size of frustrated loops has no upper limit.
To study systematically the effect of frustration, we consider now
a modified version of the model,
in which only loops up to some
specified length are considered frustrated, while longer ones
are permitted. The partition function of this model is given by
\begin{equation}
Z={\sum_C}^{(\lambda_0)} e^{\beta\mu n(C)}.
\label{eq_partition2}
\end{equation}
Here the parameters $\beta$ and $\mu$ have the same meaning as in
Eq. (\ref{eq_partition}), $n(C)$ is the number of bonds in the configuration
$C$, and the sum $\sum_C^{(\lambda_0)}$ is extended over
the bond configurations that do not contain frustrated loops of length
$\lambda\leq\lambda_0$. Thus the model switches continuously from the random
bond percolation ($\lambda_0=0$), and the $q$-bond frustrated percolation
with $q=1$ ($\lambda_0=\infty$).
\par
We have studied the critical properties of the model, for $\lambda_0=4$,
and its dynamical behavior above and below the percolation transition.
The critical exponents extracted
from the finite size scaling data collapse (see Fig. \ref{fig_statics_age4})
coincides within the errors with those of the random bond percolation,
as shown in Tab. \ref{tab_critical}.
The percolation temperature is $T_p=1.277$, intermediate between
that of the random bond percolation, $T_p=1.443$, and that of the
$q=1$ bond frustrated percolation, $T_p=1.067$.
\par
The dynamics, on the other hand, can feel the local constraint constituted
by the frustration, as the autocorrelation functions do not decay
as a single exponential when the temperature is lowered below
the percolation threshold. However, the long time regime of the relaxation
functions could be fitted with an exponential for all the temperatures
considered.
\par
In Fig. \ref{fig_relax_age4} we show the bond density autocorrelation
functions, evaluated on a lattice $L=64$, for temperatures
$T=1.277,0.911,0.625$.
The function calculated for $T=1.701$ is not shown because it overlaps
with the function calculated for $T=1.277$.
Solid curves show fits made by a pure exponential.
Averages were taken over 32 different runs,
each one taking $10^4$ steps for thermalization, and between
$2\times 10^5$ and $3\times 10^5$ steps for acquisition.
\section{Conclusions}
We have studied
the fully frustrated Ising model and the fully frustrated percolation model.
The dynamics of the models was analysed in detail, in connection
with the problem of the onset of stretched exponentials in frustrated
systems, like glasses and spin glasses.
\par
Due to absence of disorder in these models, the arguments suggested by Randeria
{\em et al.}, which predict a non-exponential relaxation below the critical
temperature $T_c$ of the corresponding ferromagnetic model,
does not apply. In fact our results show no sign of complex dynamical
behavior at $T_c$. We find instead an exponential relaxation above
the percolation temperature $T\geq T_p$, while for $T<T_p$ the long
time tail of the relaxation functions can be fitted with a stretched
exponential. So we conclude that at least in the models without disorder
the appearance of complex dynamics is related to a percolation
transition. In systems like spin glasses $T_p$ and $T_c$ are very close,
and it is difficult to distinguish numerically where the onset of
non exponential relaxation occurs.
\par
We also find that frustration plays an important role in the
critical properties at the percolation threshold $T_p$. For example, for $q=1$
the critical behavior is in the same universality of the
ferromagnetic $q=1/2$-state Potts model, contrary the unfrustrated case,
which corresponds to random bond percolation and is in the same universality
class of the $q=1$ ferromagnetic Potts model.
\par
We have  also considered a model, the locally frustrated bond
percolation, in which only loops up to length four are considered frustrated.
The model shows the same critical properties as the random bond percolation,
showing that the frustration is ``too local'' to change the universality
class. Similarly,
the relaxation functions in the long time
regime can always be fitted  with an exponential, showing that
the frustration constraint
is not enough to give rise to stretched exponential relaxation.
More careful study of this model,
possibly varying the ``range'' of the frustration between the
size of the single {\em plaquette},  and that of the whole system,
may shed more light on the role played by the frustration in the dynamics
of complex systems.
%
%
%
%%%%%%%%%%%%%%%%   REFERENCES  %%%%%%%%%%%%%%%%%%%%%%%%%%%%%%%%%%%%%%%%
%

%
%%%%%%%%%%%%%%%%%%%%%%%%%%%  FIGURES  %%%%%%%%%%%%%%%%%%%%%%%%%%%%%%
%
%%%%%%%%%%%%%%%%% fig. 1
\begin{figure}
\caption{Distribution of interactions for the fully frustrated model.
Straight lines and wavy lines correspond, respectively, to
$\epsilon_{ij}=1$ and $\epsilon_{ij}=-1$.}
\label{fig_fully}
\end{figure}
%
%%%%%%%%%%%%%%% fig. 2
\begin{figure}
\caption{Relaxation functions $f(t)$ of energy in function of time
$t$ for the fully frustrated Ising model, with spin flip dynamics,
lattice size $L=64$,
for temperatures (from left to right) T=11.0,4.5,2.269,1.701,1.110.}
\label{fig_relax_sflip}
\end{figure}
%
%%%%%%%%%%%%%%%%% fig. 3
\begin{figure}
\caption{Stretching exponents $\beta(T)$ in function of $T/T_p$, the ratio
of temperature over percolation temperature,
for the fully frustrated Ising model, with spin flip dynamics,
lattice size $L=64$.}
\label{fig_sflip_betas}
\end{figure}
%
%%%%%%%%%%%%%%%%% fig. 4a,4b
\begin{figure}
\caption{Finite size scaling of (a) $P_\infty(T)$ and (b) $\chi(T)$,
for the $q=1$ model, and for lattice sizes $L=32,48,64$.}
\label{fig_statics_q1}
\end{figure}
%
%%%%%%%%%%%%%%%%% fig. 5a,5b
\begin{figure}
\caption{Finite size scaling of (a) $P_\infty(T)$ and (b) $\chi(T)$,
for the $q=2$ model, and for lattice sizes $L=32,48,64$.}
\label{fig_statics_q2}
\end{figure}
%
%
%%%%%%%%%%%%%%%%% fig. 6
\begin{figure}
\caption{Relaxation functions $f(t)$ of bond density in function of time
$t$ for $q=1$, lattice size $L=64$,
for temperatures (from left to right) T=1.440,1.067,0.801,0.625.}
\label{fig_relax_1}
\end{figure}
%
%
%%%%%%%%%%%%%%%%% fig. 7
\begin{figure}
\caption{Stretching exponents $\beta(T)$ in function of $T/T_p$, the ratio
of temperature over percolation temperature, for the $q=1$ fully frustrated
bond percolation model, lattice size $L=64$}
\label{fig_beta_1}
\end{figure}
%
%
%%%%%%%%%%%%%%%%% fig. 8
\begin{figure}
\caption{Relaxation functions $f(t)$ of bond density in function
of time $t$ for $q=2$, lattice size $L=64$,
for temperatures (from left to right) T=2.269,1.701,1.440,1.110.}
\label{fig_relax_2}
\end{figure}
%
%
%%%%%%%%%%%%%%%%% fig. 9
\begin{figure}
\caption{Stretching exponents $\beta(T)$ in function of $T/T_p$, the ratio
of temperature over percolation temperature, for the $q=2$ fully frustrated
bond percolation model, lattice size $L=64$}
\label{fig_beta_2}
\end{figure}
%
%
%%%%%%%%%%%%%%%% fig. 10a,10b
\begin{figure}
\caption{Finite size scaling of (a) $P_\infty(T)$ and (b) $\chi(T)$,
for the local frustrated model, and for lattice sizes $L=32,48,64$.}
\label{fig_statics_age4}
\end{figure}
%
%
%%%%%%%%%%%%%%% fig. 11
\begin{figure}
\caption{Relaxation functions $f(t)$ of bond density in function of time
$t$ for the local frustrated model, lattice size $L=64$,
for temperatures (from left to right) T=1.277,0.911,0.625.}
\label{fig_relax_age4}
\end{figure}
%
%%%%%%%%%%%%%%%%%%%%%%%%%% TABLES %%%%%%%%%%%%%%%%%%%%%%%%%%%%%%%%%%
%
%
%
\begin{table}
\begin{tabular}{|c|ccc|}
model &  $1/\nu$ & $\gamma$ & $T_p$  \\
\hline
$q=1$                 & $0.56\pm 0.02$ &
                        $3.22\pm 0.07$ & $1.067\pm 0.001$ \\
$q=2$                 & $0.75\pm 0.03$ &
                        $2.34\pm 0.06$ & $1.701\pm 0.001$ \\
local ($\lambda_0=4$) & $0.75\pm 0.03$ &
                        $2.33\pm 0.04$ & $1.277\pm 0.001$ \\
\end{tabular}
\caption{Critical exponents $1/\nu$ and $\gamma$, and
percolation transition temperature $T_p$ of the fully frustrated
$q$-bond percolation model with $q=1,2$, and of the locally frustrated
bond percolation model with $\lambda_0=4$.}
\label{tab_critical}
\end{table}
\begin{table}
\begin{tabular}{|c|ccc|}
model &  $1/\nu$ & $\gamma$ & $T_c$  \\
\hline
$q=1/2$  & 0.56 & 3.27 & 1.233 \\
$q=1$    & 0.75 & 2.39 & 1.443 \\
$q=2$    & 1    & 1.75 & 2.269 \\
\end{tabular}
\caption{Critical exponents $1/\nu$ and $\gamma$, and
critical temperature $T_c$ of the ferromagnetic Potts model,
with multiplicity of spins $q=1/2,1,2$.}
\label{tab_potts}
\end{table}
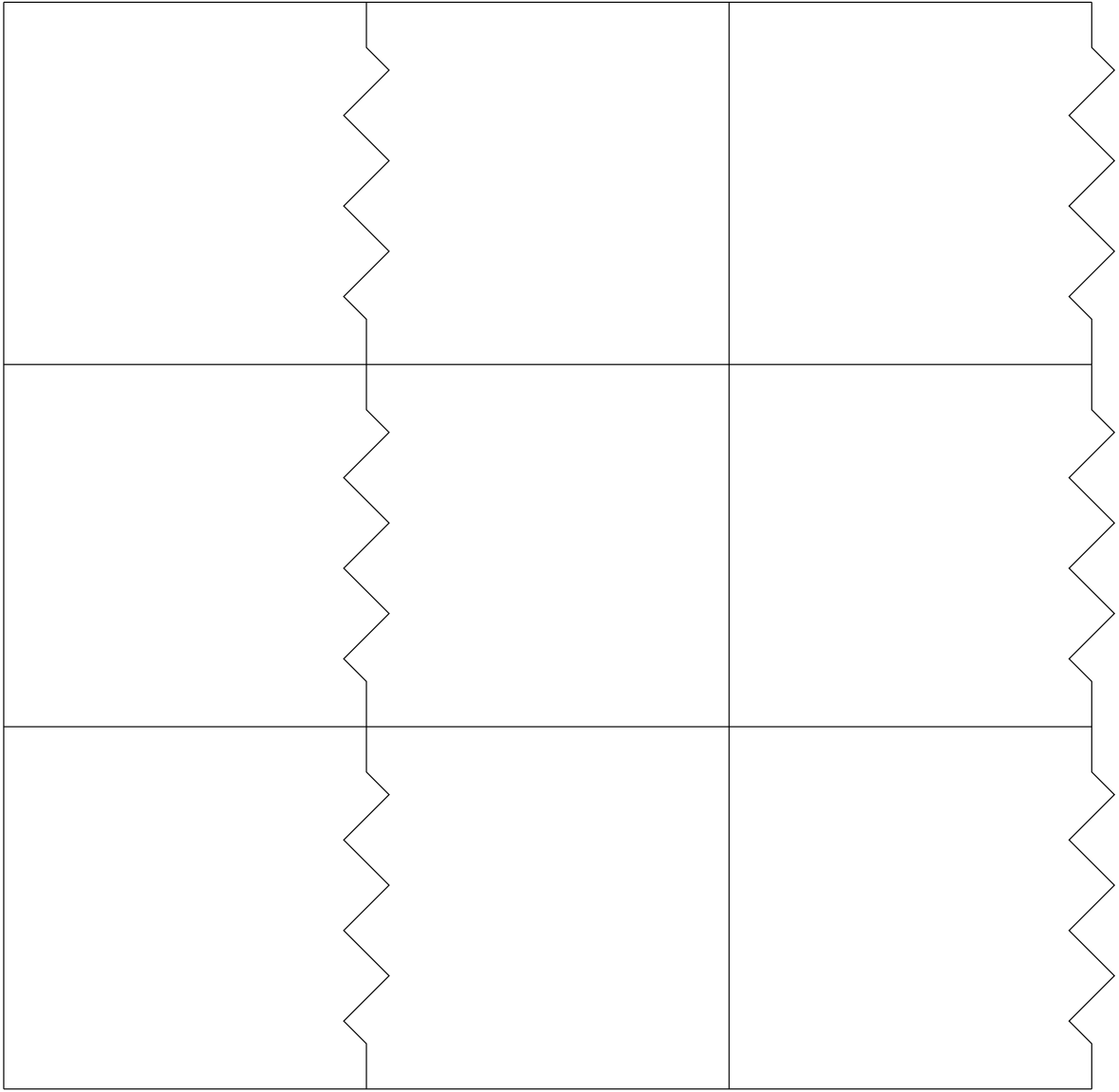
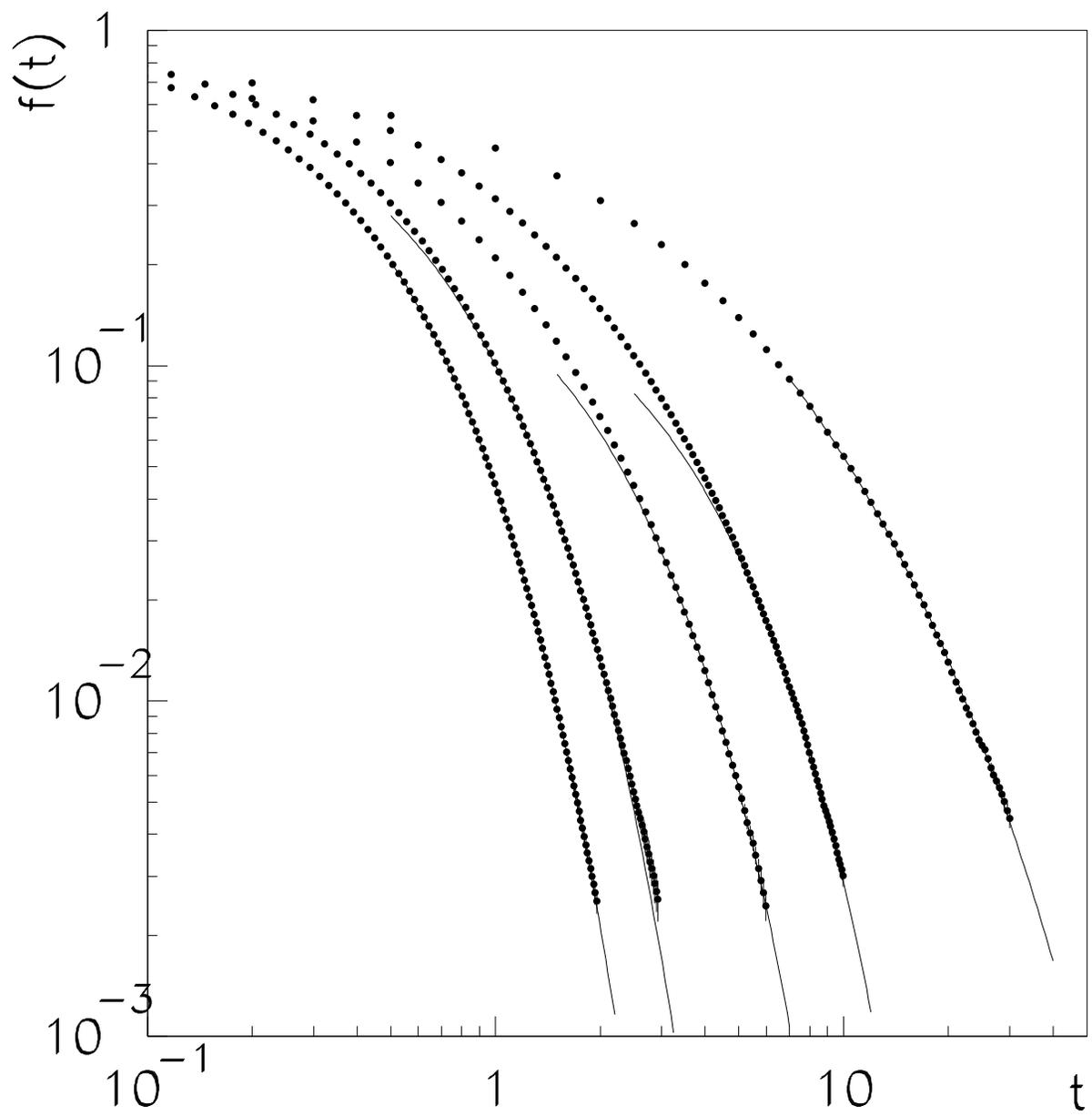
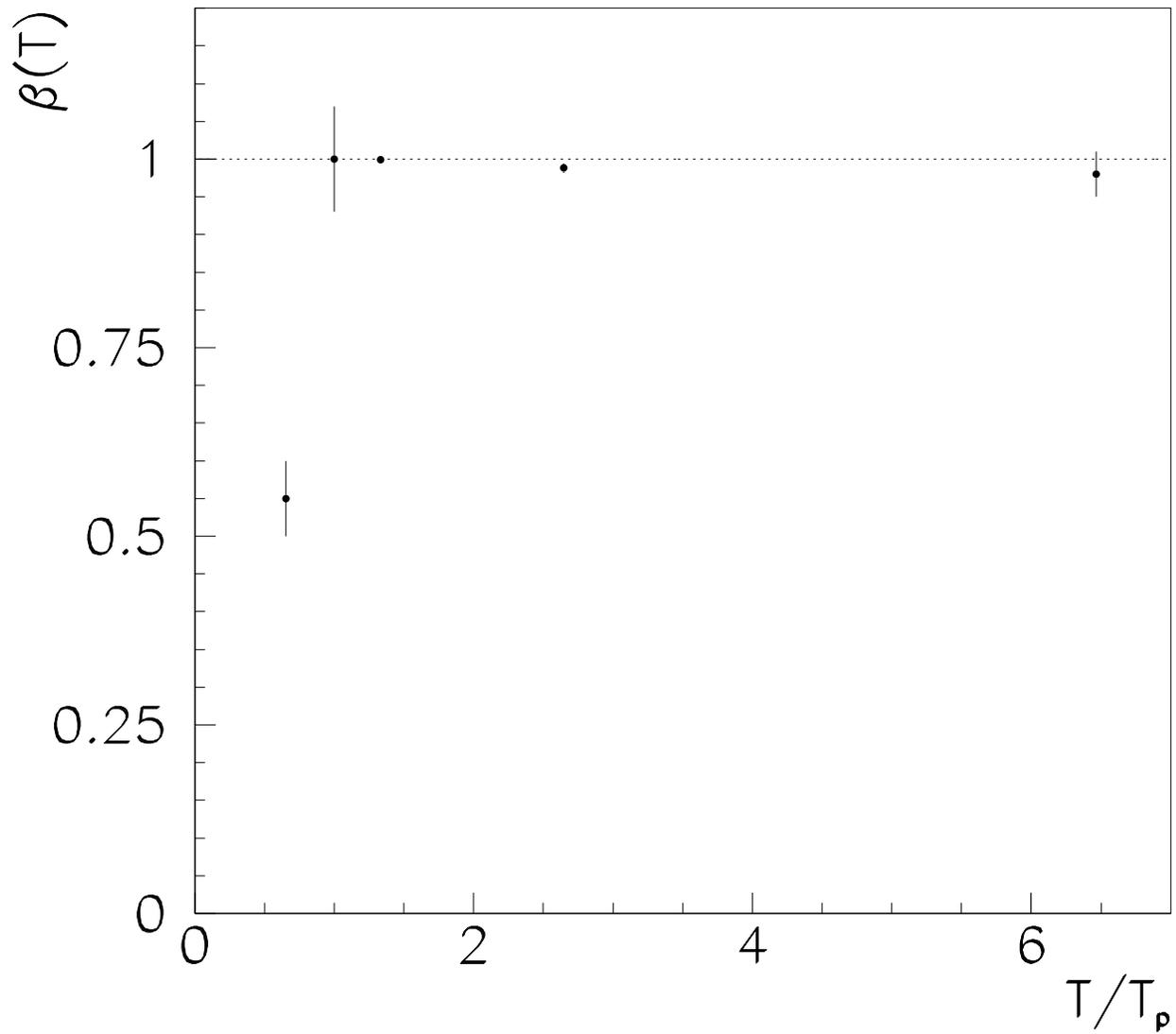
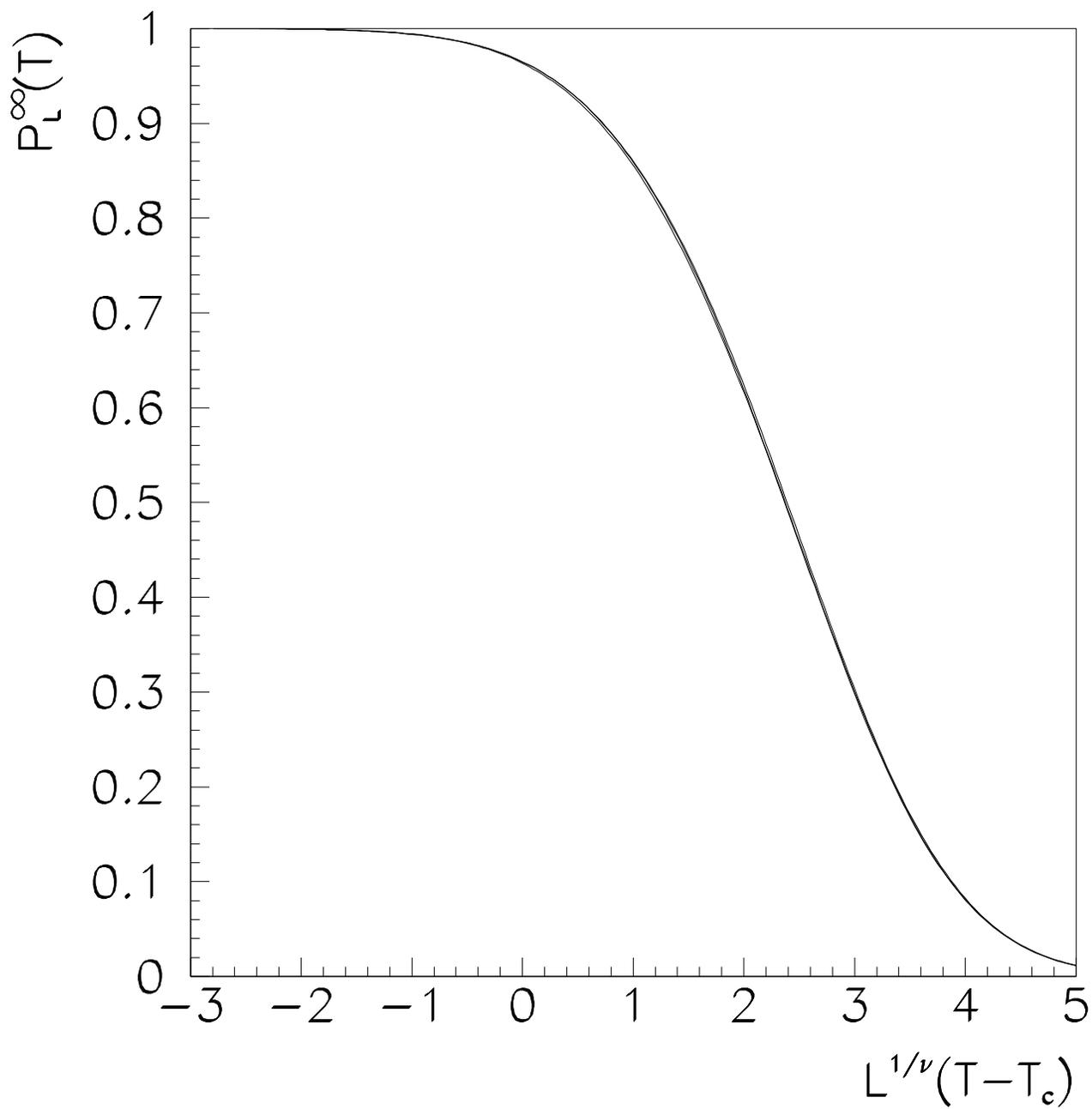
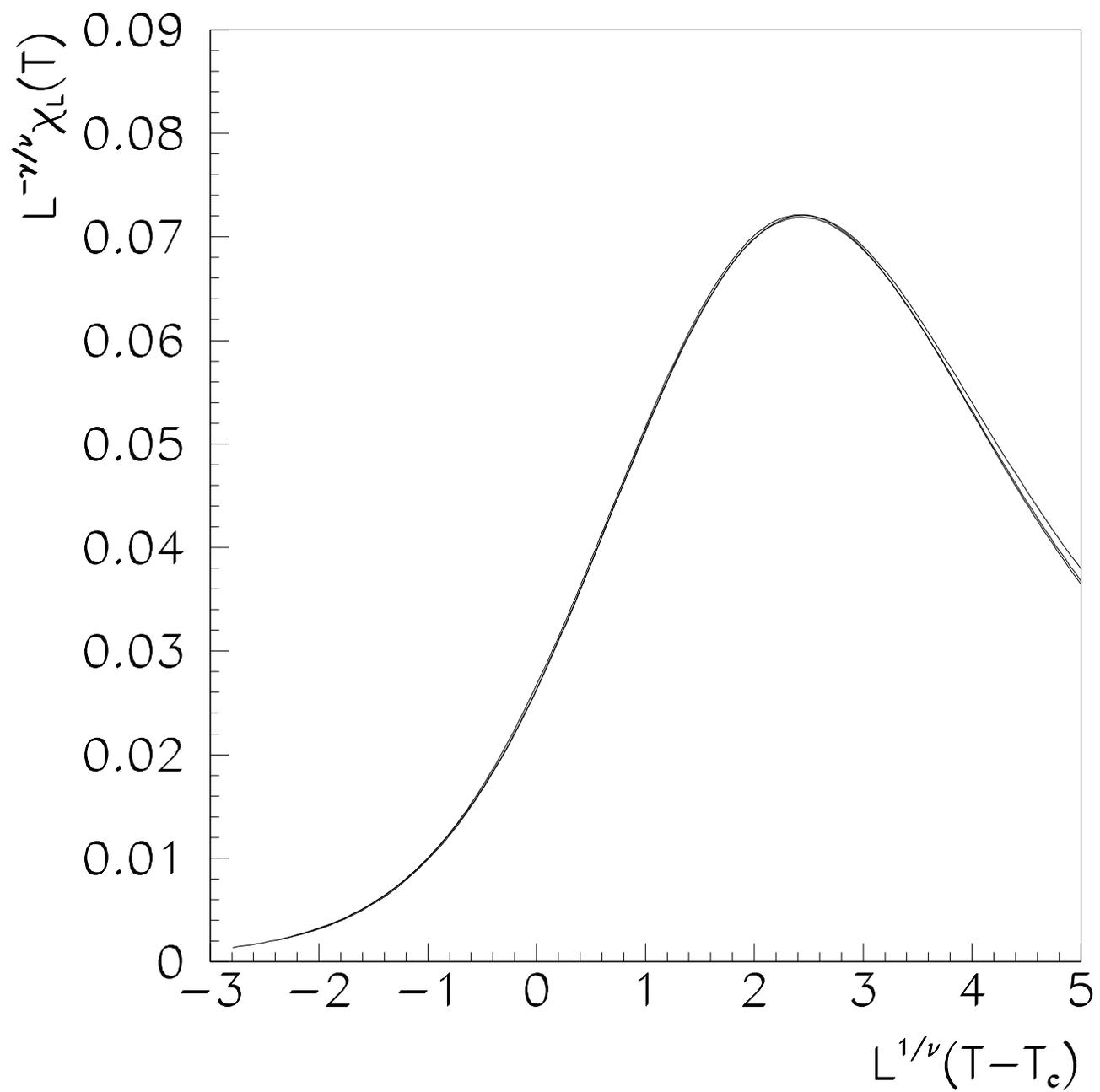
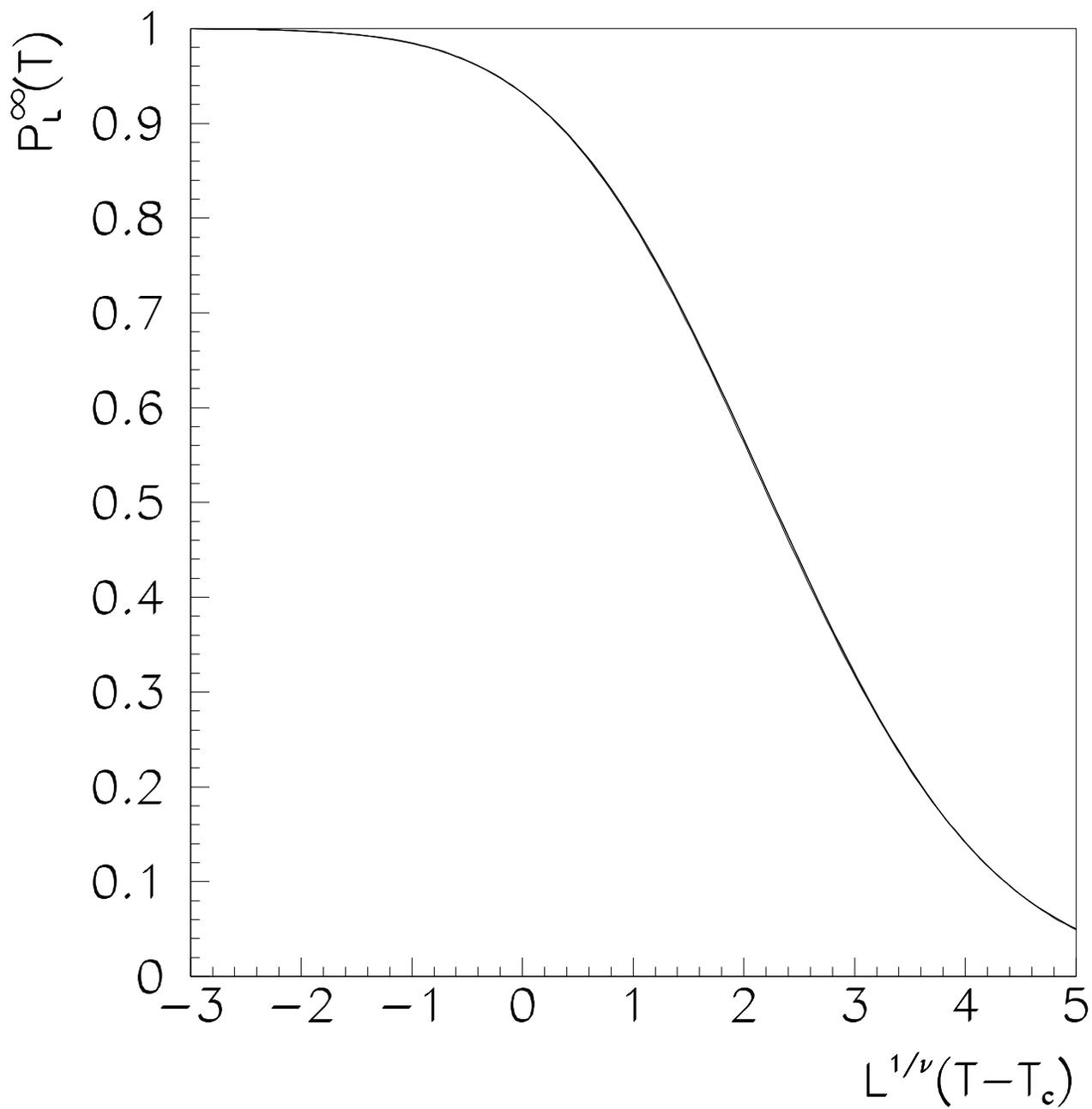
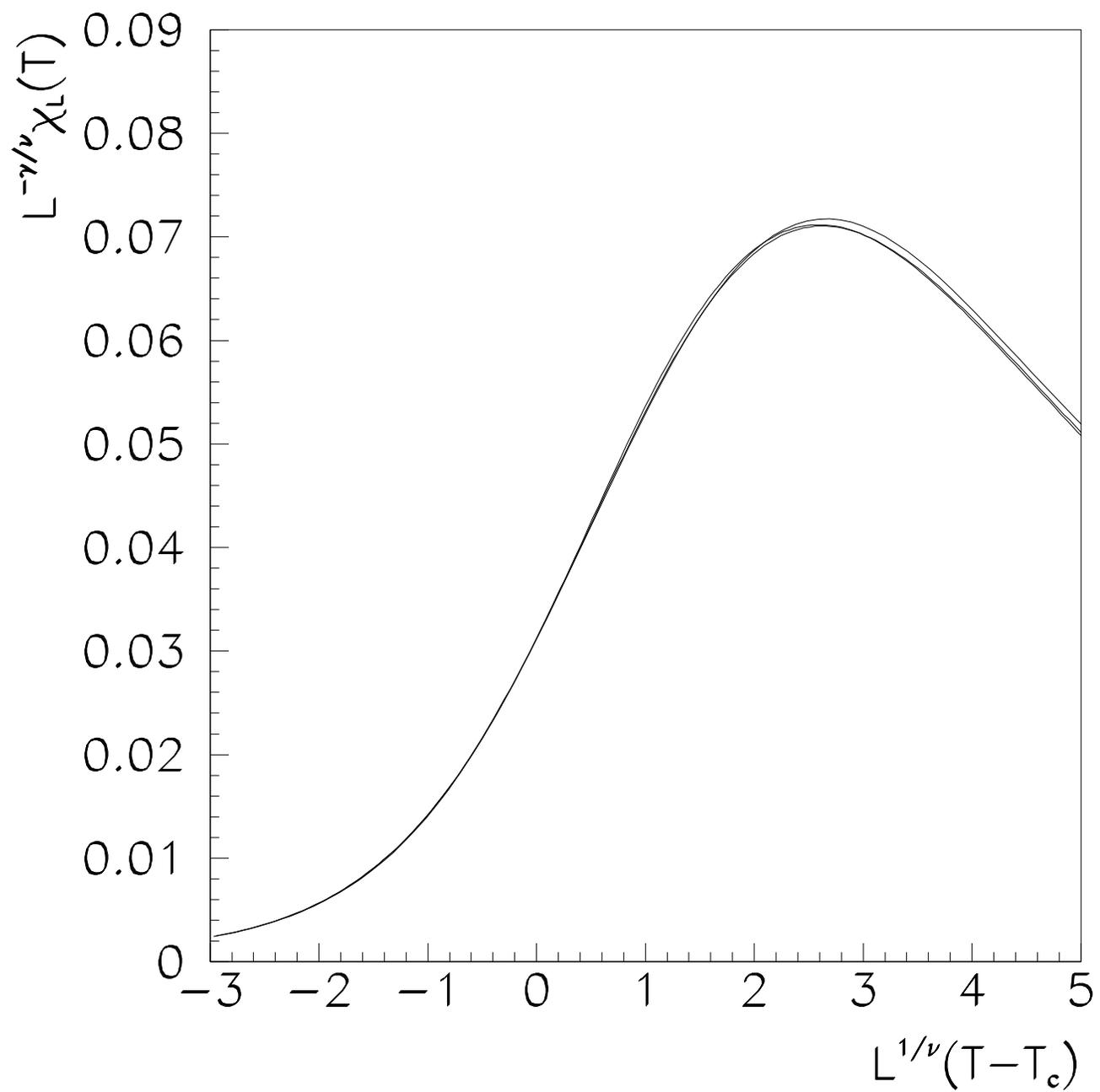
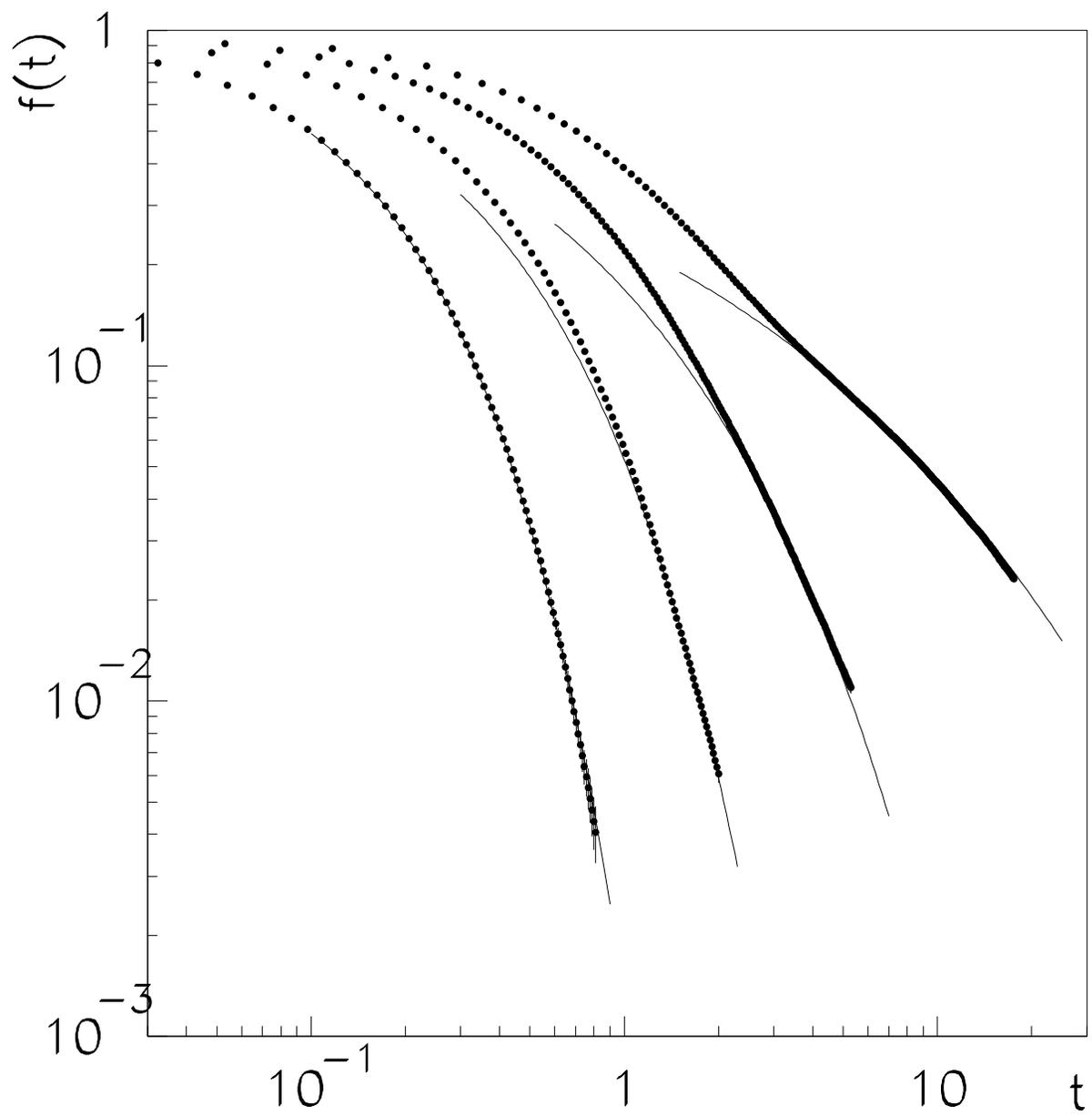
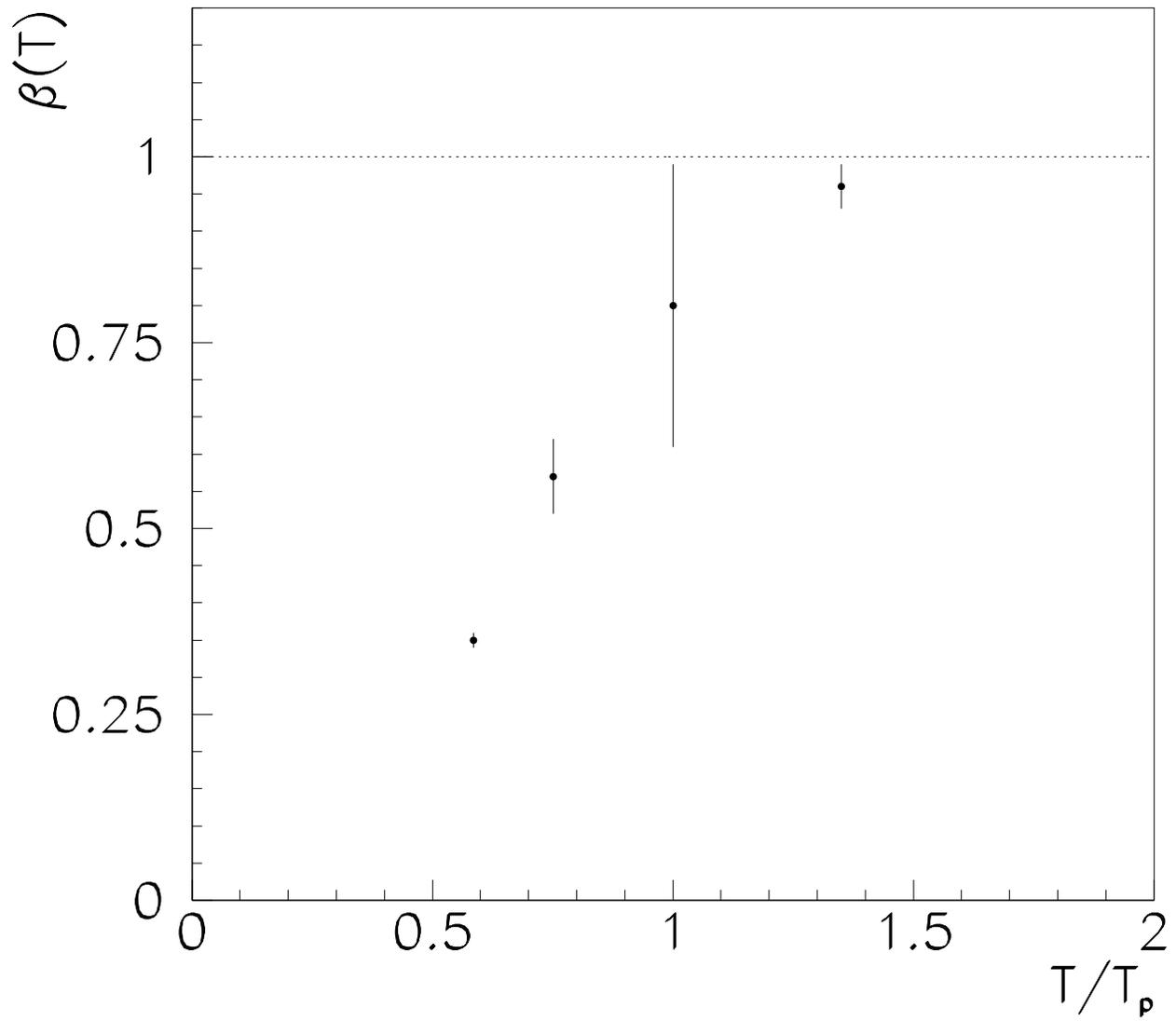
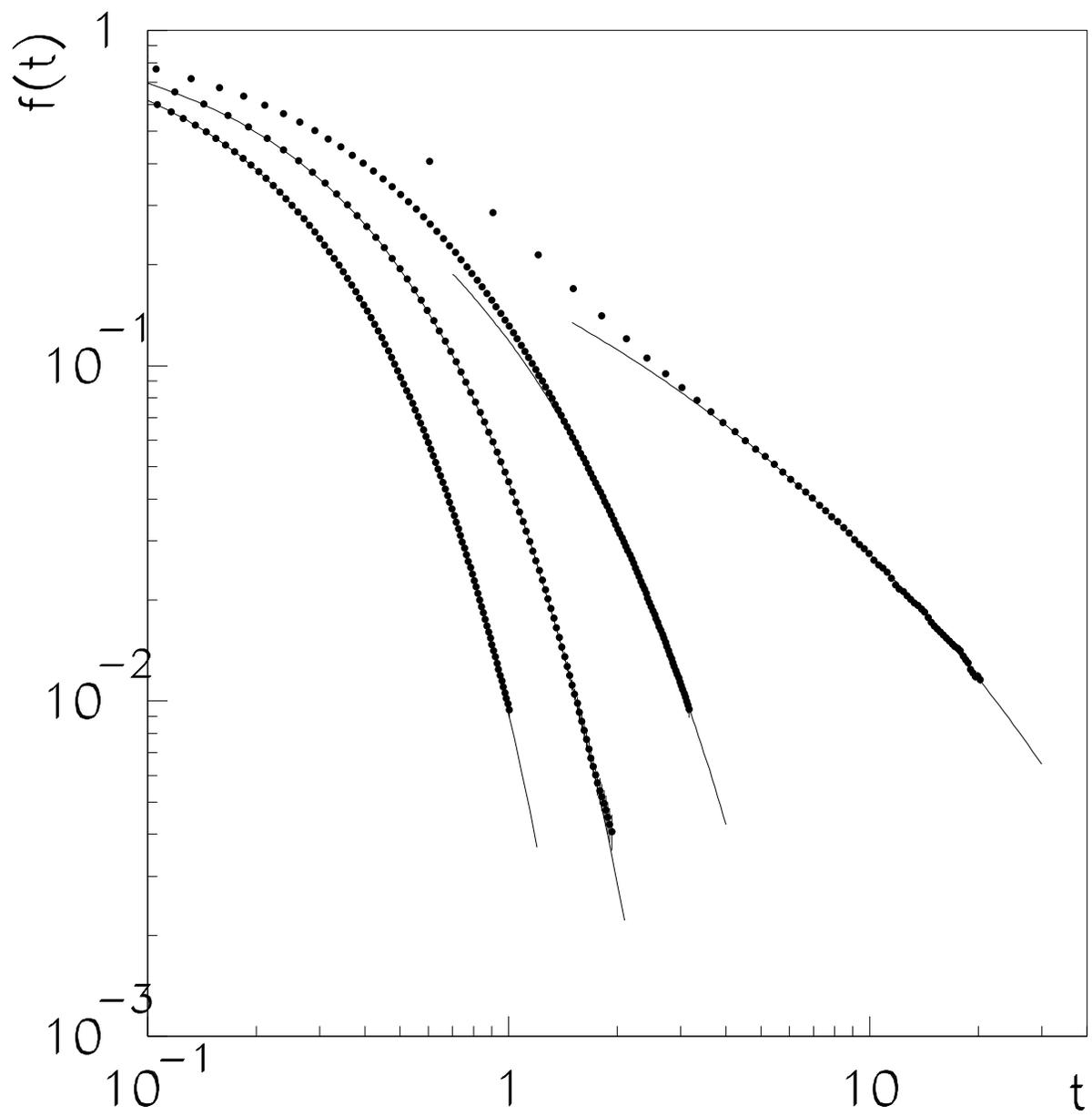
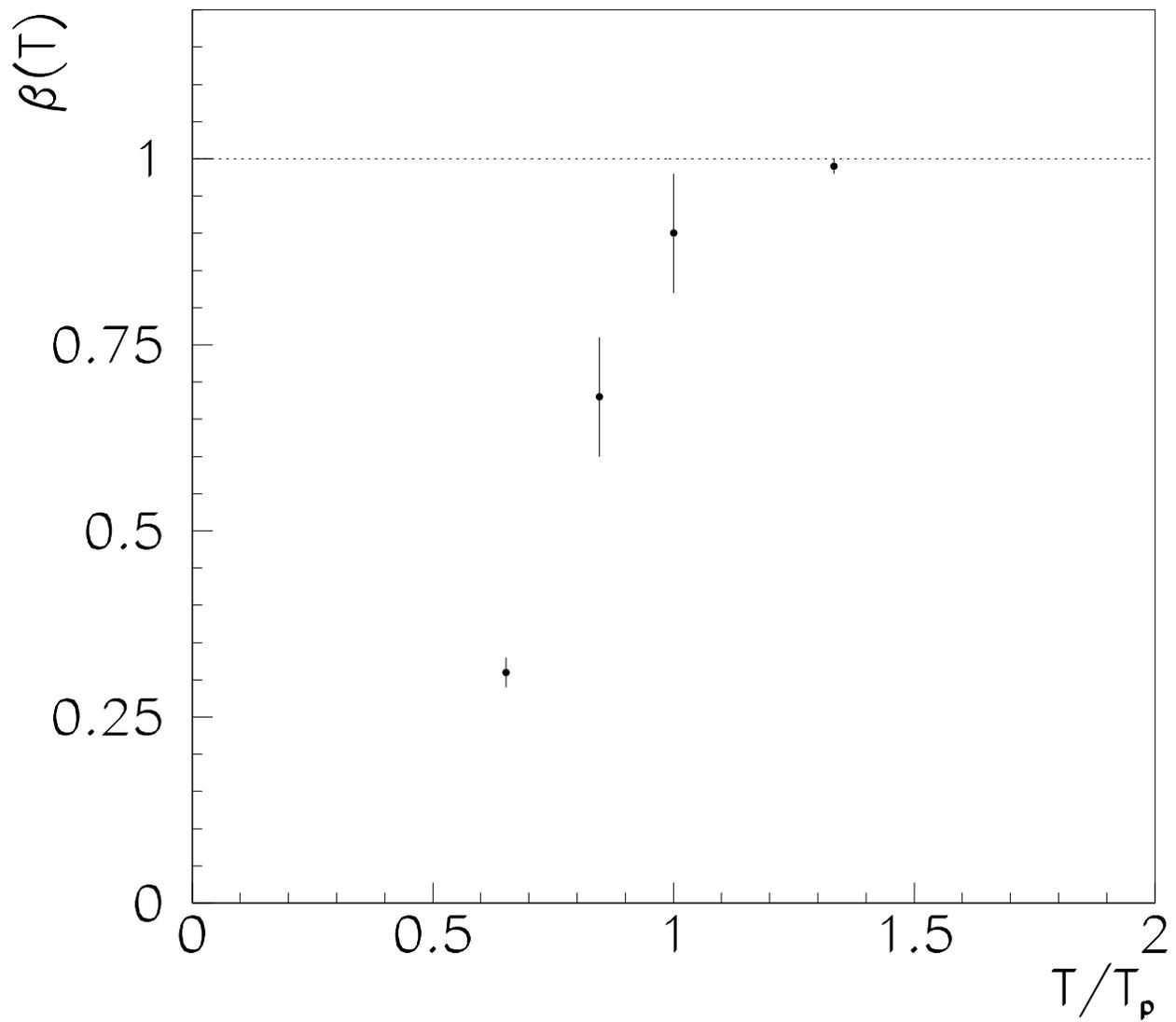
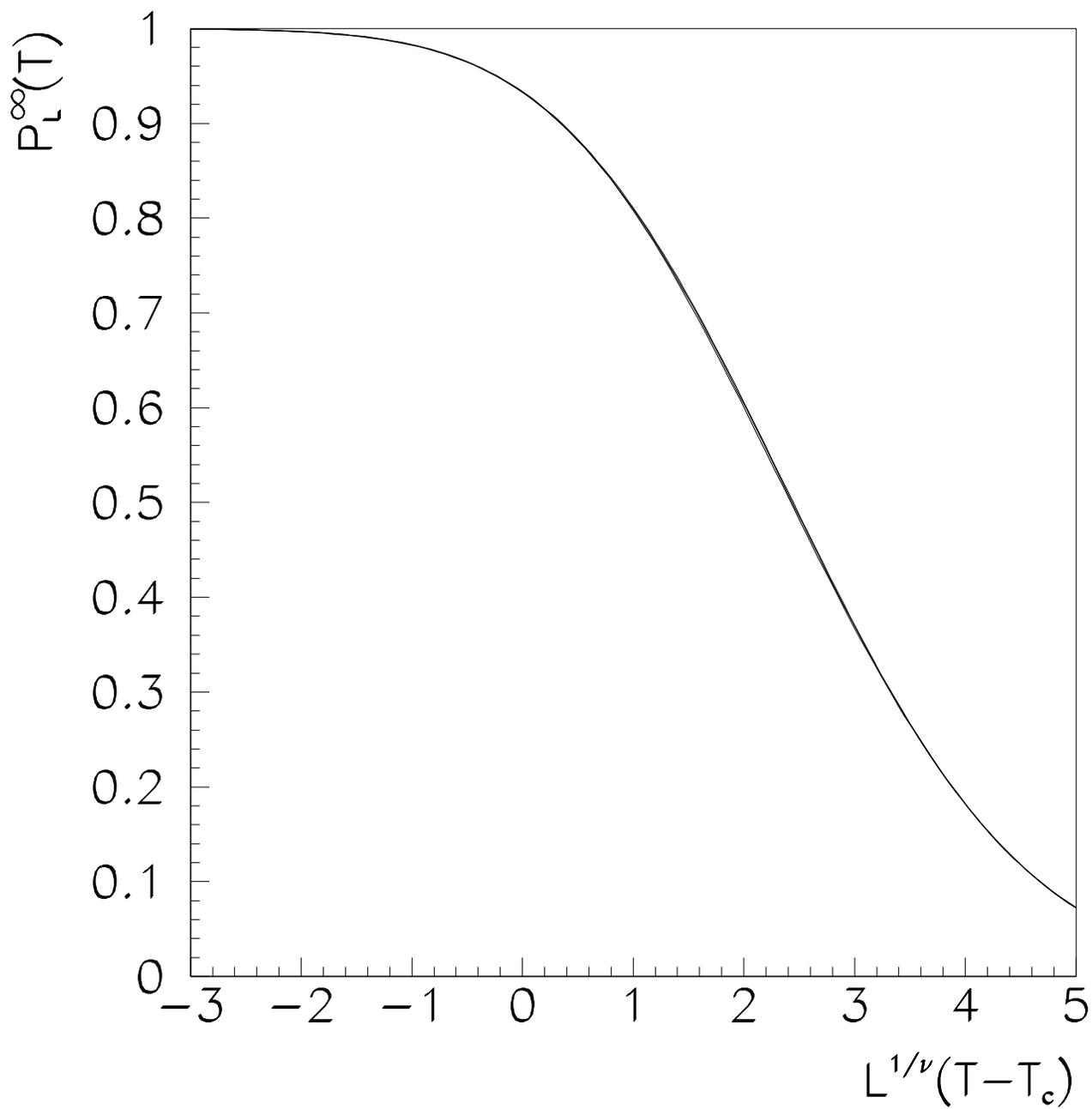
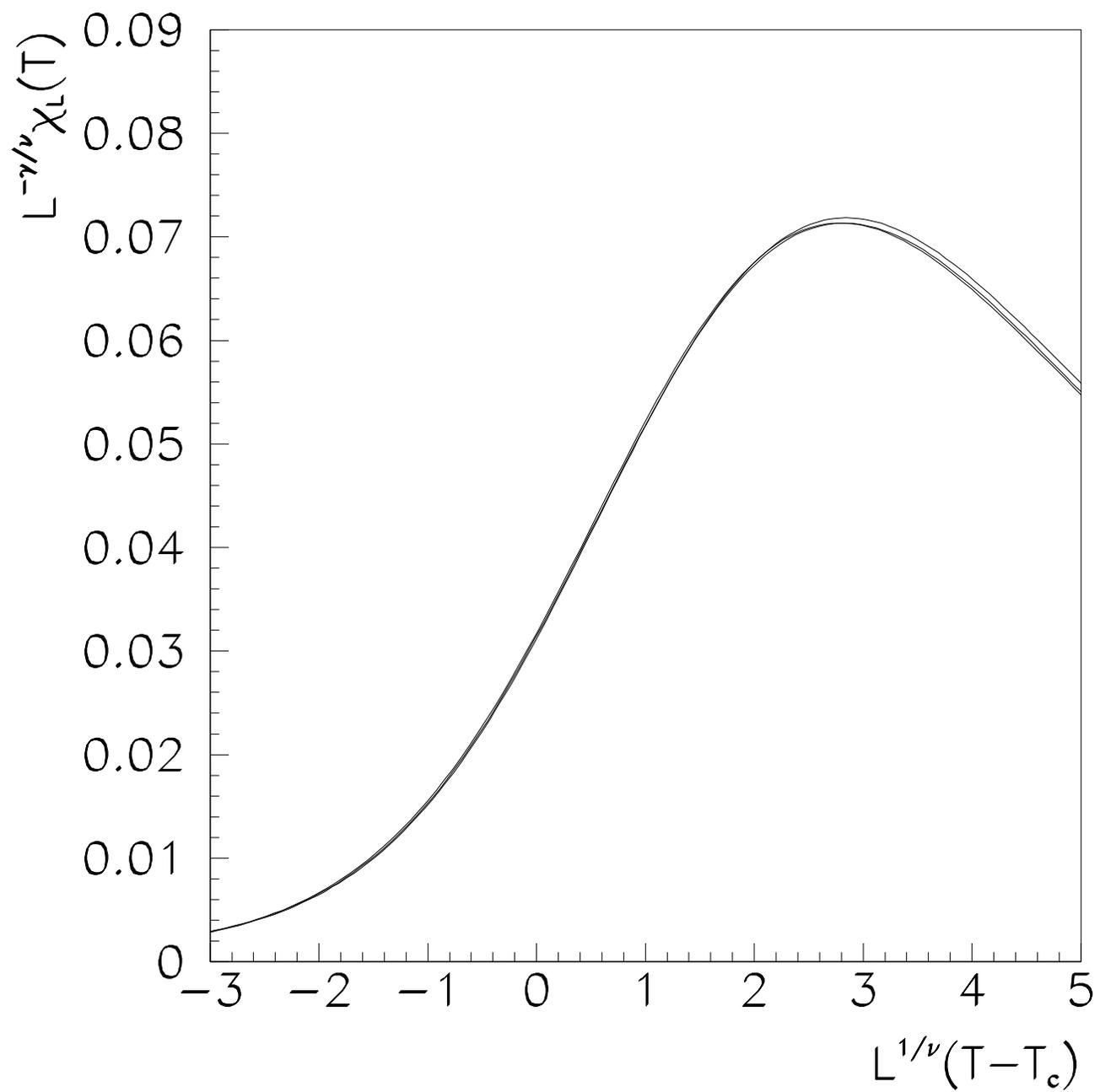
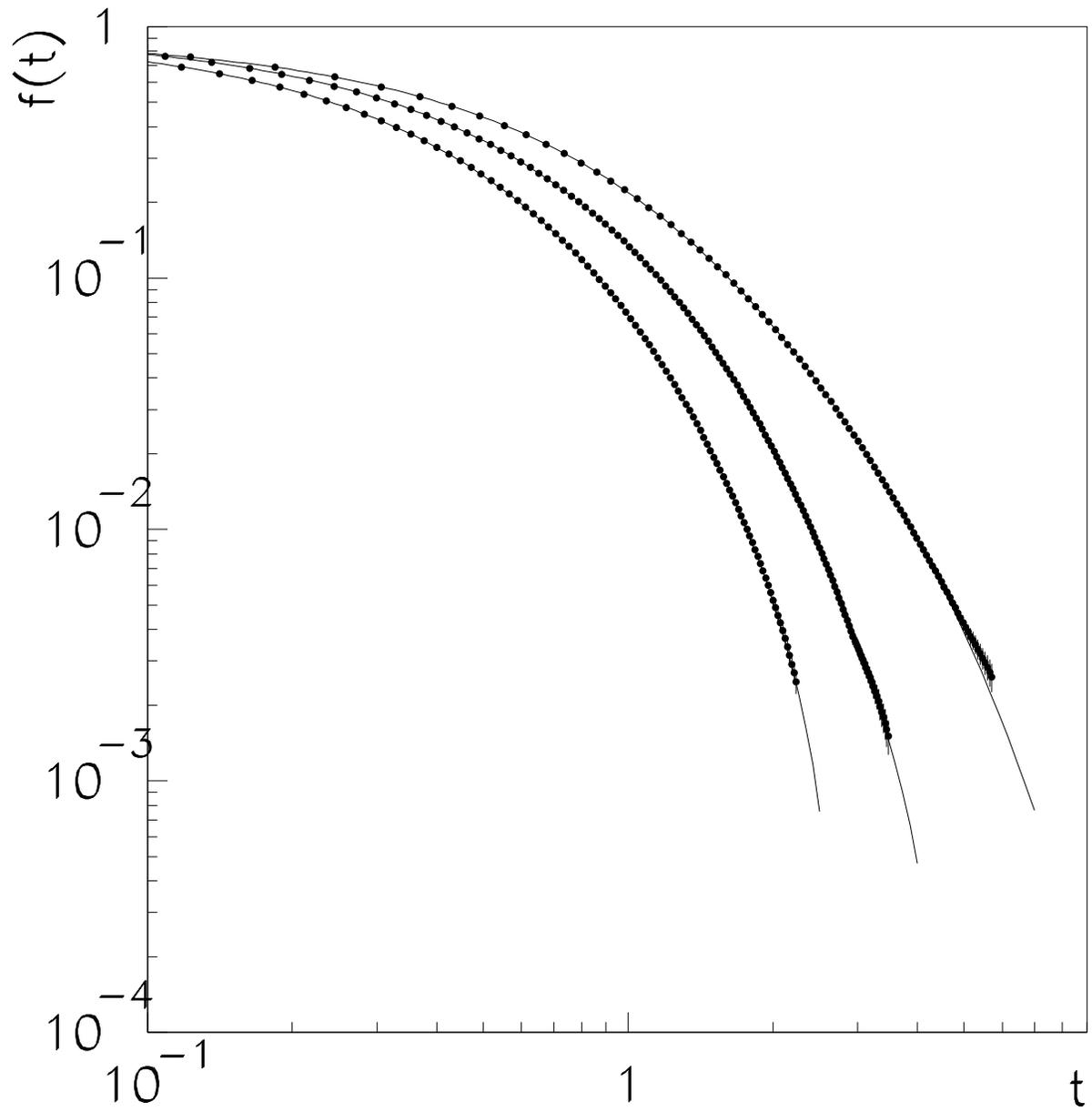
\end{document}